\documentstyle[11pt,aaspp4,psfig]{article}

\def\pp{\par\parshape 2 0truecm 15.5truecm 1truecm 14.5truecm\noindent}

\newcommand{\gtsima}{$\; \buildrel > \over \sim \;$}
\newcommand{\ltsima}{$\; \buildrel < \over \sim \;$}
\newcommand{\simgt}{\lower.5ex\hbox{\gtsima}}
\newcommand{\simlt}{\lower.5ex\hbox{\ltsima}}
\newcommand{\himpc}{{\hbox {$h^{-1}$}{\rm Mpc}} }

\begin{document}
\begin{minipage}[c]{3cm}
  \psfig{figure=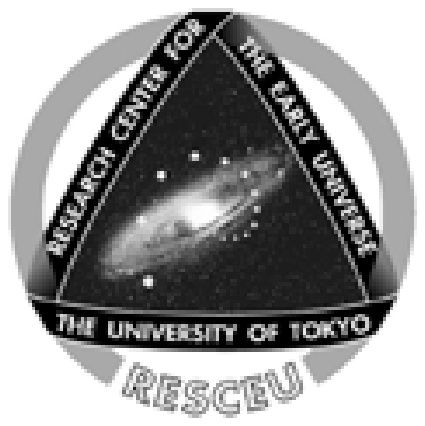,height=3cm}
\end{minipage}
\begin{minipage}[c]{9cm}
\begin{centering}
{
\vskip 0.1in
{\large \sf 
THE UNIVERSITY OF TOKYO\\
\vskip 0.1in
Research Center for the Early Universe}\\
}
\end{centering}
\end{minipage}
\begin{minipage}[c]{3cm}
\vspace{1.5cm}
RESCEU-34/97\\
UTAP-271/97
\end{minipage}\\
\vspace{0.5cm}

\title{X-ray gas density profile of clusters of galaxies \\
from the universal dark matter halo}

\bigskip

\author{Nobuyoshi Makino$^1$, Shin Sasaki$^2$, and Yasushi
Suto$^{3,4}$} 

\bigskip
\bigskip

\affil{\altaffilmark{1}  Department of Physics, 
  Ritsumeikan University, Kusatsu, Shiga, 525-77, Japan}
\affil{\altaffilmark{2}  Department of Physics, 
  Tokyo Metropolitan University, Hachioji, Tokyo 192-03, Japan}
\affil{\altaffilmark{3} Department of Physics, University of Tokyo,
  Tokyo 113, Japan} 
\affil{\altaffilmark{4} Research Center for the Early Universe
  (RESCEU), School of Science \\
  University of Tokyo, Tokyo 113, Japan} 

\bigskip

\affil{\footnotesize e-mail: makino@bkc.ritsumei.ac.jp, 
sasaki@phys.metro-u.ac.jp, suto@phys.s.u-tokyo.ac.jp}

\bigskip

\received{1997 September 10}
\accepted{1997 October 29}

\begin{abstract}
  The X-ray cluster gas density distribution in hydrostatic
  equilibrium is computed from the universal density profile of the
  dark matter halo recently proposed by Navarro, Frenk and White
  (1996, 1997). If one assumes the isothermality, the resulting
  distribution is well approximated by the conventional $\beta$-model.
  We predict the core radius $r_{\rm c}$, the $\beta$-parameter, and
  the X-ray luminosity of clusters as a function of the temperature
  $T_{\rm X}$ of clusters in some representative cosmological models,
  and compare them with observations and results of numerical
  simulations.  The predicted size of $r_{\rm c}$ is a factor of ($3
  \sim 10$) smaller than the average of observed values.  If both the
  universal density profile and the hydrostatic equilibrium are
  reasonable approximation to the truth, then this suggests either
  that the previous X-ray observations systematically overestimate the
  core radius of gas densities in clusters of galaxies, or that some
  important physical mechanisms, which significantly increase the core
  radius, is still missing.
\end{abstract}


\vfill \centerline{\sl The Astrophysical Journal, Part 1, in press}


\section{Introduction}

While it has been recognized for a long time that the gas density
profile of X-ray clusters is well approximated by the isothermal
$\beta$-model:
\begin{equation}
n_g(r) = {n_{g0} \over [1+(r/r_c)^2]^{3\beta/2}} ,
\label{eq:betaprofile}
\end{equation}
the origin of this functional form is not yet accounted for. Navarro,
Frenk, \& White (1996, 1997; NFW96 and NFW97 hereafter) found that the
virialized halos in numerical simulations are well fitted by the
following universal profile:
\begin{equation}
\rho_{\rm DM}(r) = {\delta_c \rho_{c0} \over
(r/r_s)(1+r/r_s)^2 } ,
\label{eq:nfwprofile}
\end{equation}
where $\rho_{c0}$ is the critical density of the universe at $z=0$,
and
\begin{eqnarray}
\label{eq:deltac}
\delta_c(M) &\approx& 3\times 10^3 \Omega_0 [1+z_{\rm f}(M)]^3, \\
\label{eq:rs}
r_s(M) &=& {r_{\rm vir}(M) \over c(M)} 
= {1 \over c(M)} \left({3M \over 4\pi\Delta_c\rho_{c0}} \right)^{1/3}.
\end{eqnarray}
In the above expressions, $\Omega_0$ is the density parameter at
$z=0$, $\Delta_c(\Omega_0,\lambda_0)$ is the collapse factor in a
spherical nonlinear model, $z_{\rm f}(M)$ is the average formation
redshift of objects of mass $M$, and the concentration parameter
$c(M)$ is related to $\delta_c(M)$ as
\begin{equation}
\delta_c = {\Delta_c \over 3} {c^3 \over \ln(1+c) -c/(1+c)} .
\end{equation}

NFW96 also mentioned that under the gravitational potential of the
dark halo (\ref{eq:nfwprofile}), the gas in hydrostatic equilibrium
reaches a profile quite similar to equation (\ref{eq:betaprofile}).
Using a hydro-dynamical simulation, Eke, Navarro, \& Frenk (1997, ENF
hereafter) examined in more details the properties of X-ray clusters
in a cold dark matter model with $\Omega_0=0.3$, $\lambda_0=0.7$,
$h=0.7$ and $\sigma_8=1.05$, where $\lambda_0$ is the dimensionless
cosmological constant, $h$ is the Hubble constant in units of
100km$\cdot$sec$^{-1}\cdot$Mpc$^{-1}$, and $\sigma_8$ is the top-hat
mass fluctuation amplitude at $8\himpc$. They also found that
isothermal $\beta$-model describes well the simulated cluster gas
distribution.

In the present {\it Letter} we derive an analytical expression for the
gas density profile embedded in the universal dark matter halo
(\ref{eq:nfwprofile}) assuming hydrostatic equilibrium and isothermal
distribution (\S 2).  The resulting distribution is remarkably well
fitted by equation (\ref{eq:betaprofile}). With this we are able to
predict the core radius, the $\beta$-parameter, and the X-ray
luminosity of clusters as a function of the mass $M$ or temperature
$T_{\rm X}$ of clusters once the cosmological model is specified. Our
predictions are compared with the results of ENF simulations and
observations in \S 3.  Finally \S 4 is devoted to discussion of
further implications.

\section{Hydrostatic equilibrium of gas and dark matter}

Consider an isothermal spherical gas cloud with temperature $T_{\rm
X}$, then its density distribution $\rho_g$ in hydrostatic equilibrium
satisfies
\begin{equation}
  {kT_{\rm X} \over \mu m_p}{d \ln \rho_g \over dr} = - {G M(r) \over r^2} ,
\label{eq:equilib}
\end{equation}
where $\mu$ and $m_p$ denote the mean molecular weight (we adopt 0.59
below) and the proton mass.  If one neglects the gas and galaxy
contributions to the gravitational mass in the right-hand side, then
equation (\ref{eq:nfwprofile}) yields that
\begin{equation}
M(r) = 4\pi \delta_c\rho_{c0} r_s^3 
\left[\ln\left(1+{r\over r_s}\right)-{r\over r+r_s}\right] ,
\label{eq:mhalo}
\end{equation}
and equation (\ref{eq:equilib}) can be analytically integrated to give
\begin{equation}
\rho_{\rm g}(r) = \rho_{\rm g0}~\exp\left[-{27 \over 2}b
\left(1-{\ln(1+r/r_s) \over r/r_s}\right)\right] 
= \rho_{\rm g0}~e^{-27b/2}~(1+r/r_s)^{27b/(2r/r_s)} ,
\label{eq:gasprofile}
\end{equation}
with 
\begin{equation}
b(M)\equiv {8\pi G\mu m_p\delta_c(M)\rho_{c0}r_s^2 \over 27kT_{\rm X}}.
\label{eq:bdef}
\end{equation}

The cluster gas temperature $T_{\rm X}$ is expected to be close to the
virial temperature $T_{\rm vir}(M)$ of the dark halo.  In the profile
(\ref{eq:nfwprofile}), the latter is in fact dependent on the radius
$r$:
\begin{equation}
kT_{\rm vir}(r) = \gamma {G\mu m_p M(r) \over 3 r},
\label{eq:tvir}
\end{equation}
where $\gamma$ is a fudge factor of order unity which should be
determined by the efficiency of the shock heating of the gas; ENF
adopted $\gamma=1.5$ while Kitayama \& Suto (1997) adopted 1.2 as
their canonical value in the analysis of X-ray cluster number
counts. If one substitutes equation (\ref{eq:tvir}) into equation
(\ref{eq:bdef}), one finds that
\begin{equation}
b(r) = {2 \over 9\gamma}{r \over r_s} 
\left[\ln\left(1+{r \over r_s}\right)- {r \over r+r_s}\right]^{-1} .
\end{equation}
Throughout the present paper we assume that the cluster gas is
isothermal; this is observationally true for most clusters, and
theoretically expected due to the effect of the thermal conduction.
In that case it is reasonable to adopt either $T_{\rm vir}(r_s)$ or
$T_{\rm vir}(r_{\rm vir})$ as the gas temperature $T_{\rm X}$.  This
corresponds to $b(r_s) \approx 1.15/\gamma$, or $b(r_{\rm
vir})=2c/(9\gamma)[\ln(1+c)-c/(1+c)]^{-1} \approx (1.3\sim2)/\gamma$ .

\begin{figure}
\begin{center}
  \leavevmode\psfig{file=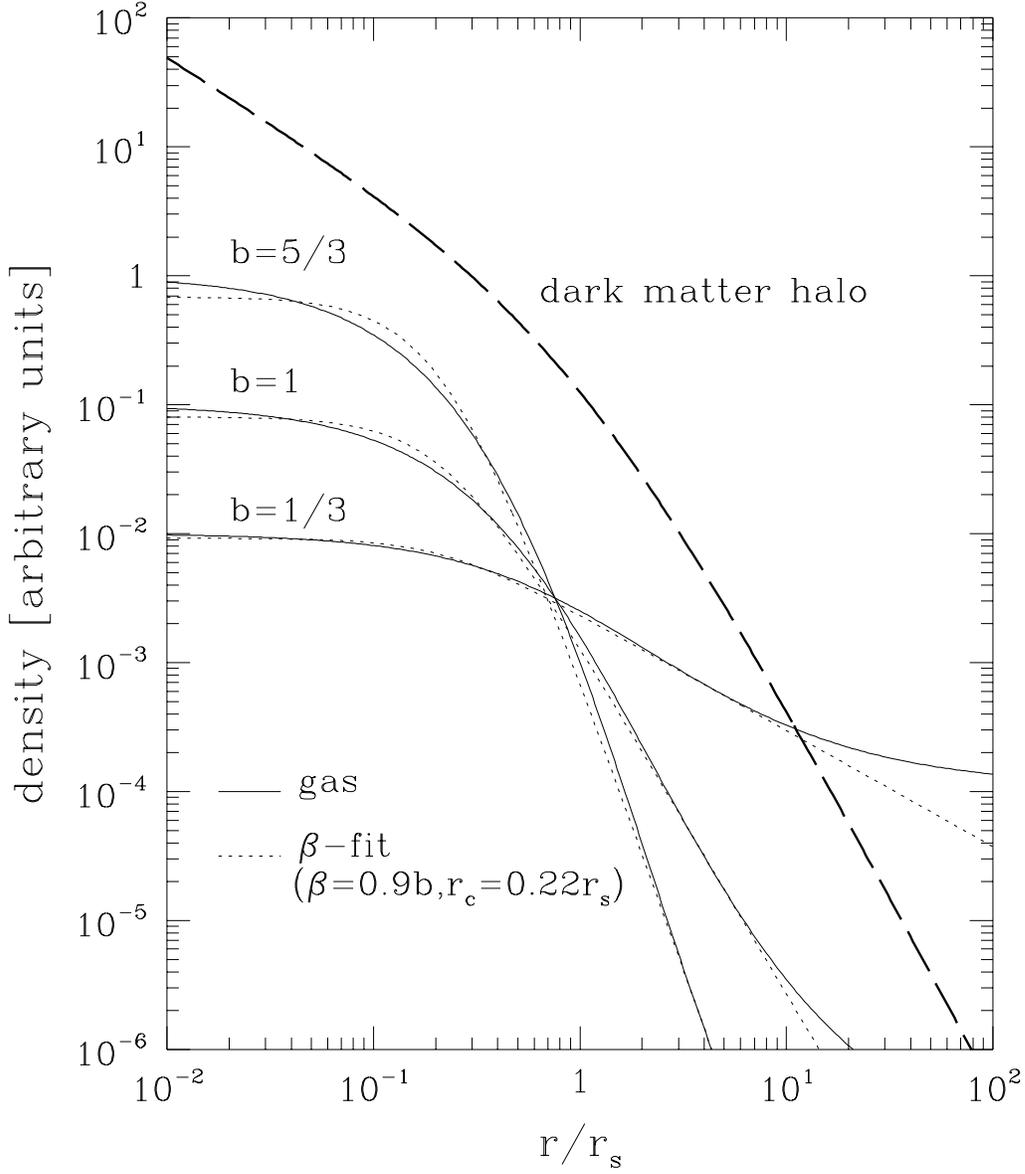,height=18cm}
\end{center}
\caption{Gas density profile (solid lines) 
expected from the universal density profile of dark matter halo
(dashed line) for $b=1/3$, $1$, and $5/3$. For comparison, the
best-fit $\beta$-models with $\beta=0.9b$ and $r_c=0.22r_s$ are
plotted in dotted lines.
\label{fig:densprof}}
\end{figure}

The derived functional shape (\ref{eq:gasprofile}) may appear
completely different from equation (\ref{eq:betaprofile}) at a first
glance.  As is shown in Figure \ref{fig:densprof}, however, it is
surprisingly well approximated by the isothermal $\beta$-model
profile:
\begin{equation}
\rho_{\rm g}(r) = {\rho_{\rm g0} A(b)\over 
[1+(r/r_{\rm c,eff})^2]^{3\beta_{\rm eff}/2}} ,
\label{eq:betafit}
\end{equation}
with $A(b)=-0.178b+0.982$, $r_{\rm c,eff}= 0.22r_s$ and $\beta_{\rm
eff}=0.9b$ over $0.01r_s<r<10r_s$.  Interestingly this implies that
$\beta_{\rm eff}(r_s) \approx 1/\gamma$ and $ \beta_{\rm eff}(r_{\rm
vir})(1\sim2)/\gamma$ if we take $T_{\rm X}=T_{\rm vir}(r_s)$ and
$T_{\rm vir}(r_{\rm vir})$, respectively (also see Fig.\ref{fig:rc} b
below). With $\gamma= 1 \sim 1.5$, those values agree very well with
typical observed ones, i.e., $\beta = 1.2 \sim 0.6$.  This is why
Waxman \& Miralda-Escud\'{e} (1995), NFW96 and ENF were able to
reproduce a gas distribution very similar to the $\beta$-model in
their specific models, and in fact the agreement should be quite
generic and almost independent of the adopted parameters.

\section{Implications of the derived gas density profile}

Given the apparent success in describing the observed shape of the
X-ray cluster gas profiles in the universal dark matter halo
potential, it is reasonable to examine the predicted scales of the
physical variables in further details. In fact, all the relevant
quantities can be computed using Appendix A of NFW97 once the
cosmological model is fully specified.

Figure \ref{fig:rc} plots the ratio of $r_{\rm vir}$ and $r_{\rm
c,eff} (= 0.22r_s)$ ({\it panel a}), the ratio of the virial
temperatures at $r_s$ and at $r_{\rm vir}$ ({\it panel b}), and
$r_{\rm c,eff}$ ({\it panel c}), against the temperature $T_{\rm X}
\equiv T_{\rm vir}(r_s)$. In what follows we set $\gamma=1.5$ for
definiteness, but it does not change the conclusions below.  We adopt
the shape parameter $\Gamma$ to characterize the power spectrum, which
is equivalent to $\Omega_0 h \exp[-\Omega_b(1+\sqrt{2h}/\Omega_0)]$ in
cold dark matter (CDM) models with the baryon density parameter
$\Omega_b$ (Sugiyama 1995).  The representative cosmological models
which we consider include SCDM ($\Omega_0=1$, $\lambda_0=0$,
$\Gamma=0.5$, $\sigma_8=1.2$), OCDM ($\Omega_0=0.3$, $\lambda_0=0$,
$\Gamma=0.25$, $\sigma_8=1.0$), and LCDM ($\Omega_0=0.3$,
$\lambda_0=0.7$, $\Gamma=0.2$, $\sigma_8=1.0$).

Although $r_{\rm vir}/r_c$ is around 50 on scales of rich clusters
(Fig.\ref{fig:rc}a), the corresponding temperatures $T_{\rm
vir}(r_{\rm vir})$ and $T_{\rm vir}(r_s)$ agree within a factor of two
(Fig.\ref{fig:rc}b). Therefore the assumption of isothermal clusters
is acceptable, and one obtains reasonable values for $\beta_{\rm eff}$
comparable to the observed one (\S 2) even if we adopt either $T_{\rm
vir}(r_{\rm vir})$ or $T_{\rm vir}(r_s)$ as the gas temperature.  In
Figure \ref{fig:rc}c, we also plot the $r_c$ and $T_{\rm X}$ relations
of observed clusters from the {\it Einstein} (Jones \& Forman 1984;
David et al. 1993), and {\it EXOSAT} (Edge \& Stewart 1991a,b; see
also Kitayama \& Suto 1996) observations.  Clearly the predicted
$r_{\rm c,eff}$ is a factor of ($3 \sim 10$) smaller than the average
of observed values.

Equation (\ref{eq:equilib}) does not involve the amplitude of gas
density, the central density $\rho_{\rm g0}$ should be determined
by some other consideration. For this purpose, we use
\begin{equation}
\frac{M_{\rm gas}(r_{\rm vir})}{M(r_{\rm vir})} 
=   \frac{f_{\rm gas}\Omega_b}{\Omega_0} ,
\label{eq:gasfrac}
\end{equation}
where $f_{\rm gas}$ is the gas mass fraction of the total baryon in
the cluster. Observationally $f_{\rm gas}$ is typically $0.7 \sim
0.9$. With equations (\ref{eq:mhalo}) and (\ref{eq:gasprofile}),
equation (\ref{eq:gasfrac}) reduces to
\begin{equation}
  \rho_{\rm g0} = 
\frac{f_{\rm gas}\Omega_b \rho_{\rm c0} \delta_c }{\Omega_0}
 e^{27b/2}
\left[ \ln(1+c) - {c \over 1+c} \right]
\left[ \int_0^c x^2 (1+x)^{27b/2x} dx
\right]^{-1} .
\label{eq:rhog0}
\end{equation}
This can be translated to the central number density of electron
$n_{\rm e0}$ assuming the primordial abundance of hydrogen and helium
($X=0.76$), which is plotted in Figure \ref{fig:ne0lxtx}a.  For
definiteness, we set $b=0.7/0.9$ so as to reproduce a typical shape of
observed X-ray clusters, and adopt $\Omega_b=0.015h^{-2}$ and $f_{\rm
gas}=0.8$.  As is clear from Figure \ref{fig:ne0lxtx}a, the predicted
$n_{\rm e0}$ is larger than typical observed values. This is a direct
consequence of the small $r_{\rm c}$ in this model given the total gas
mass ratio in the cluster at $r_{\rm vir}$ (Fig.\ref{fig:rc}c).

Once $n_{\rm e0}$ is specified, it is straightforward to compute the
X-ray bolometric luminosity:
\begin{eqnarray}
  L_{\rm X,bol}(T_{\rm X}) 
&=& 4\pi \int_0^{r_{\rm vir}} r^2dr~ \alpha(T_{\rm X}) n_{\rm e}^2(r)
\nonumber \\
&=&  4\pi \alpha(T_{\rm X}) n_{\rm e0}^2 r_s^3 e^{-27b}
\int_0^c x^2 (1+x)^{27b/x} dx ,
\label{eq:lxbol}
\end{eqnarray}
where $\alpha(T_{\rm X})$ denotes the bolometric emissivity for which
we include line emissions in addition to the bremsstrahlung (Masai
1984, Kitayama, Sasaki \& Suto 1998). The results are plotted in
Figure \ref{fig:ne0lxtx}b, together with the observed $L_{\rm X,bol}$ --
$T_{\rm X}$ relation (David et al. 1993; Ebeling et al. 1996; Ponman
et al. 1996). Since $L_{\rm X,bol}$ predicted from equation
(\ref{eq:lxbol}) scales as $h^{-3}$ unlike the observed ones ($\propto
h^{-2}$), we divide our predictions by $h$ assumed in each model in
Figure \ref{fig:ne0lxtx}b.  Dash-dotted line indicates the best-fit to
the observed relation
\begin{equation}
  L_{\rm X, bol} = 2.9 \times 10^{44} h^{-2}
\left(\frac{T}{6{\rm keV}}\right)^{3.4} {\rm ~ erg~sec}^{-1},
\end{equation}
(Kitayama \& Suto 1997). The shallower slope of the predicted $L_{\rm
X,bol}$ -- $T_{\rm X}$ relation ($\sim 2$) than the observed one
($\approx 3.4$) reflects a self-similar nature of the evolution in
this model (Kaiser 1986). In fact our model predictions reproduce the
results of Kitayama \& Suto (1997; their Fig.1) and of ENF (their
Fig.15), apart from the small difference due to the different choice
of adopted parameters.

\section{Discussion and conclusions}

One of the most important consequences of the universal density
profile is that all the physical quantities characterizing the gas
density profile ($r_{\rm c}$, $r_{\rm vir}$, $\beta_{\rm eff}$,
$L_{\rm X, bol}$, $n_{\rm e0}$) are computed as functions of the total
halo mass $M$, or almost equivalently of the corresponding gas
temperature $T_{\rm X}$, once underlying cosmological parameters
($\Omega_0$, $\lambda_0$, $\Gamma$, $h$, and $\sigma_8$) are fully
specified. This can be regarded as a significant theoretical
improvement of the understanding in the physical origin of the
conventional $\beta$-model.

On the other hand, if both the universal density profile and the
hydrostatic equilibrium are reasonable approximation to the truth,
then our result indicates either that the previous X-ray observations
systematically overestimate the core radius of clusters of galaxies,
or that we neglect some unknown important physical mechanisms which
significantly increases the core radius. Incidentally $r_{\rm vir}$ in
equation (\ref{eq:rs}) would be further divided by $1+z_{\rm f}$ if
$z_{\rm f}$ really corresponds to the formation redshift of the entire
cluster, and this would make $r_{\rm c}$ even smaller.

Figure \ref{fig:massprof} shows the mass profile of the universal dark
matter halo. Also plotted are the estimates on the basis of the
$\beta$-model fitting (eqs.[\ref{eq:betafit}] and [\ref{eq:equilib}]).
The mass inferred from the gas density in this way underestimates the
true value for $r<r_c$.  In turn, if the X-ray observations
overestimate the core radius of clusters of galaxies, they would
underestimate the total gravitational mass of the clusters.

\begin{figure}
\begin{center}
  \leavevmode\psfig{file=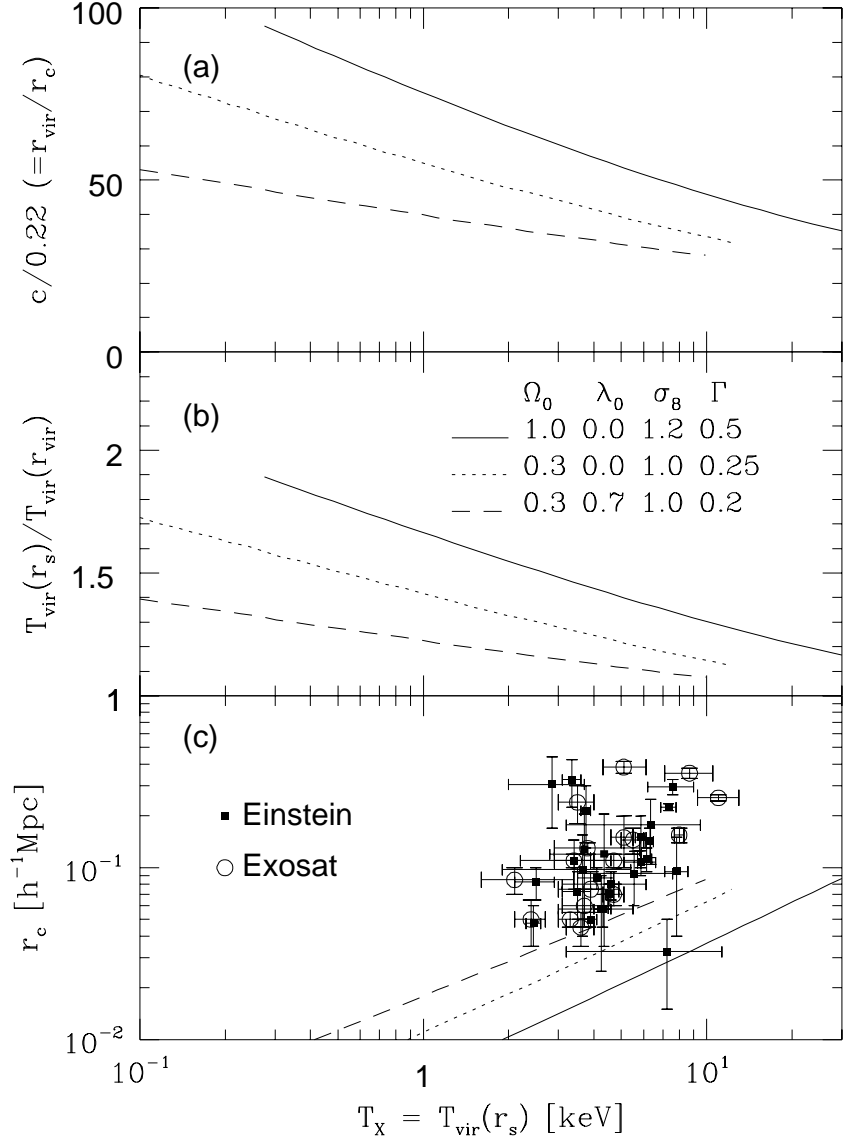,height=18cm}
\end{center}
\caption{Predicted properties of gas density distribution as functions 
of the X-ray cluster gas temperature $T_X=T_{\rm vir}(r_s)$.  (a)
Ratio of $r_{vir}$ and $r_c (\approx 0.22 r_s)$; (b) ratio of the
virial temperatures at $r_s$ and at $r_{\rm vir}$; (c) predicted sizes
of the effective core radius compared with the observed cluster data.
Solid, dotted and dashed lines indicate SCDM, OCDM, and LCDM models,
respectively.
\label{fig:rc}}
\end{figure}

\begin{figure}
\begin{center}
  \leavevmode\psfig{file=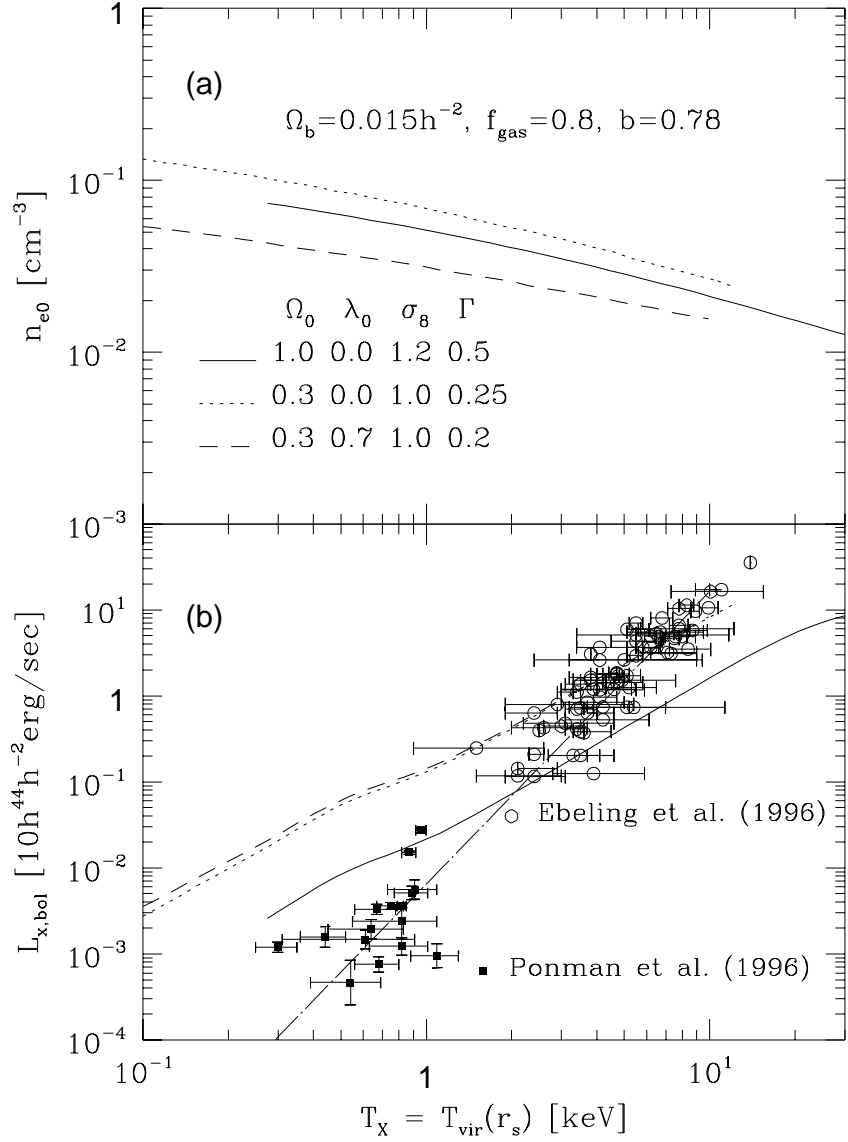,height=18cm}
\end{center}
\caption{Central electron number density and 
bolometric X-ray luminosity plotted against the gas temperature for
$\Omega_b=0.015h^{-2}$, $f_{\rm gas}=0.8$ and $b=0.78$ ($\beta_{\rm
eff}=0.7$).
\label{fig:ne0lxtx}}
\end{figure}

\begin{figure}
\begin{center}
  \leavevmode\psfig{file=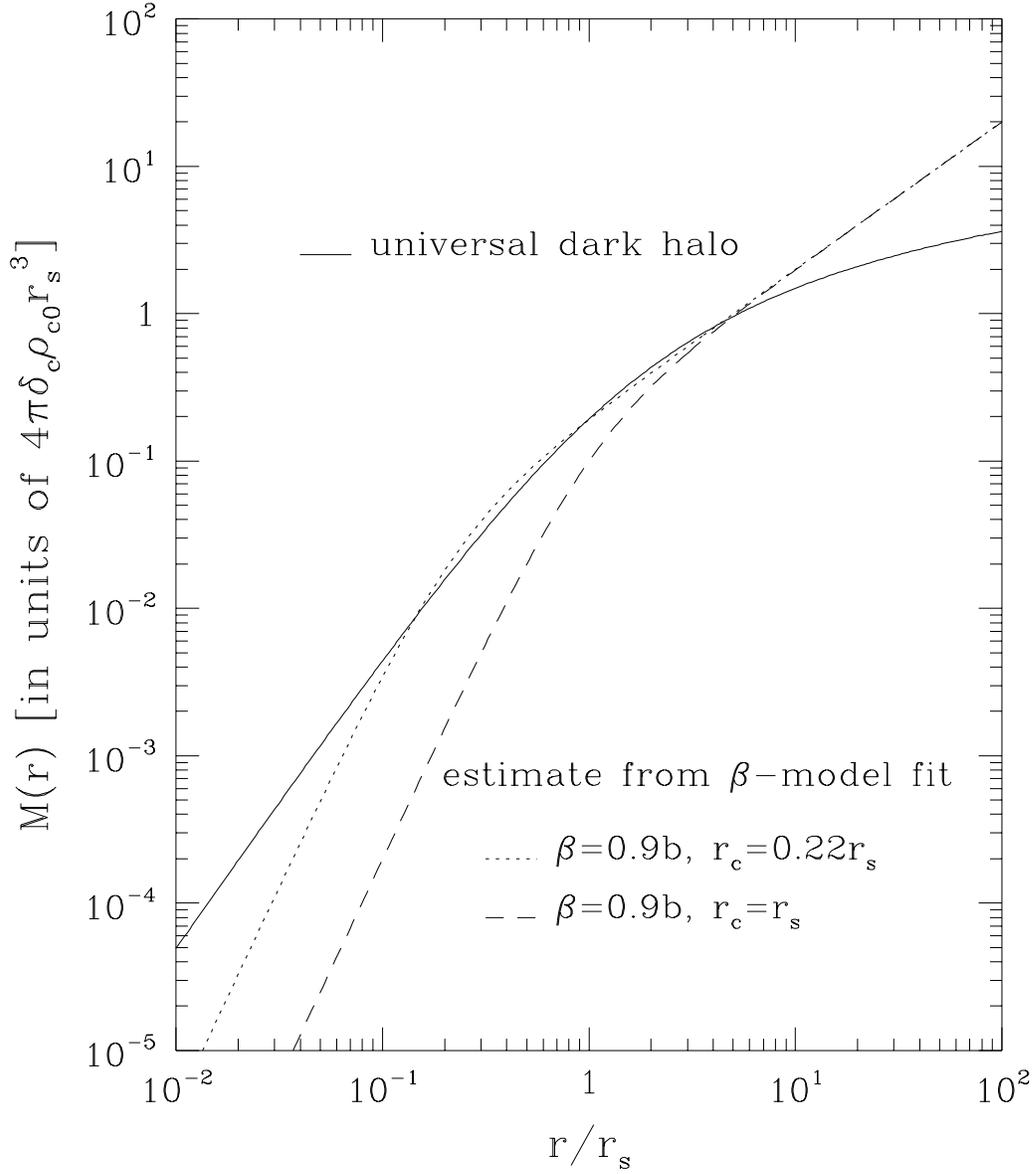,height=18cm}
\end{center}
\caption{Mass profile of the universal dark matter halo 
(solid line) compared with the estimates on the basis of the
$\beta$-model fitting. Dotted line indicates the estimate for the
best-fit $\beta$-model ($\beta=0.9b$ and $r_c=0.22r_s$), while dashed
line corresponds to $\beta=0.9b$ and $r_c=r_s$.
\label{fig:massprof}}
\end{figure}

ENF obtained that $r_{\rm vir}=2.1h^{-1}$Mpc, $r_{\rm
s}=0.3h^{-1}$Mpc, $r_{\rm c}=0.1h^{-1}$Mpc, and $n_{\rm
e0}=2\times10^{-2}h^{-2}$cm$^{-3}$ by averaging over the ten most
massive clusters at $z=0$. Their $r_{\rm vir}$, $r_{\rm s}$ and
$r_{\rm c}$ are systematically larger than, but yet are consistent
(within a factor of two) with, our LCDM predictions
(Figs. \ref{fig:rc} and \ref{fig:ne0lxtx}); this might be partly
ascribed to the insufficient number of particles to resolve the small
core, and partly to the fact that they include clusters not completely
in equilibrium.  Fukushige \& Makino (1997) and more recently Moore et
al. (1997) argued that their simulations with much higher spatial
resolution show the steeper inner density profile like $\propto
r^{-1.4}$ which should predict the core radius much smaller than what
we discussed above.

\bigskip
\bigskip

We thank Yipeng Jing for providing numerical routines to compute
$\delta_c(M,z)$ in various cosmologies.  We are grateful to Tetsu
Kitayama and Noriko Yamasaki for useful discussions, and to Tatsushi
Suginohara for comments on the manuscript. This research was supported
in part by the Grants-in-Aid for the Center-of-Excellence (COE)
Research of the Ministry of Education, Science, Sports and Culture of
Japan (07CE2002) to RESCEU (Research Center for the Early Universe),
University of Tokyo, Japan

\clearpage
\centerline{\bf REFERENCES}

\def\apjpap#1;#2;#3;#4; {\pp#1, {#2}, {#3}, #4}
\def\apjbook#1;#2;#3;#4; {\pp#1, {#2} (#3: #4)}
\def\apjppt#1;#2; {\pp#1, #2.}
\def\apjproc#1;#2;#3;#4;#5;#6; {\pp#1, {#2} #3, (#4: #5), #6}

\apjpap David, L. P., Slyz, A., Jones, C., Forman, W., \& Vrtilek, 
S. D. 1993;ApJ;412;479;  
\apjpap Ebeling, H., Voges, W., B\"{o}hringer, H., Edge, A. C.,
Huchra, J. P., \& Briel, U. G. 1996;MNRAS;281;799; 
\apjpap Edge, A. C., \& Stewart, G. C. 1991a;MNRAS;252;414;
\apjpap Edge, A. C., \& Stewart, G. C. 1991b;MNRAS;252;428;
\apjppt Eke, V.R, Navarro, J.F., \& Frenk, C.S. 1997;
      ApJ, submitted (astro-ph/9708070: ENF);
\apjpap Fukushige, T., \& Makino, J. 1997;ApJ;477;L9;
\apjpap Jones, C., \& Forman, W. 1984;ApJ;276;38;
\apjpap Kaiser, N. 1986;MNRAS;222;323;
\apjpap Kitayama, T., \& Suto, Y. 1996;ApJ;469;480;
\apjpap Kitayama, T., \& Suto, Y. 1997;ApJ;490;
  in press (astro-ph/9702017);
\apjppt Kitayama, T., Sasaki,S., \& Suto, Y. 1998;PASJ, in press
 (astro-ph/9708088);
\apjpap Masai, K. 1984;Ap\&SS;98;367;
\apjppt Moore, B., Governato, F., Quinn, T., Stadel, J., \& Lake, G.
1997;ApJ, submitted (astro-ph/9709051);
\apjpap Navarro, J.F., Frenk, C.S., \& White, S.D.M. 1996;ApJ;462;
   563 (NFW96);
\apjpap Navarro, J.F., Frenk, C.S., \& White, S.D.M. 1997;ApJ;490;
  in press (astro-ph/9611107: NFW97);
\apjpap Ponman, T. J., Bourner, P. D. J., Ebeling, H., \&
B\"{o}hringer, H. 1996;MNRAS;283;690;
\apjpap Sugiyama,N., 1995;ApJS;100;281;
\apjpap Waxman, E., \& Miralda-Escud\'{e}, J. 1995;ApJ;451;451;

\bigskip 

\end{document}